\documentclass[a4paper,preprint]{revtex4}
\usepackage{hyperref}
\usepackage[english]{babel}
\usepackage[latin1]{inputenc}
\usepackage{amsmath}
\usepackage{amssymb}
\usepackage{graphicx}
\usepackage{natbib}

\begin{document}
\title{
A global hybrid coupled model based on Atmosphere--SST feedbacks
}
\author{Andrea A. Cimatoribus}
\email{cimatori@knmi.nl}
\affiliation{Royal Netherlands Meteorological Institute, De Bilt, The Netherlands}
\author{Sybren S. Drijfhout}
\affiliation{Royal Netherlands Meteorological Institute, De Bilt, The Netherlands}
\author{Henk A. Dijkstra}
\affiliation{Institute for Marine and Atmospheric research Utrecht, Utrecht University, Utrecht, The Netherlands
}

\begin{abstract}
  A global hybrid coupled model is developed, with the aim of studying the effects of ocean--atmosphere feedbacks on the stability of the Atlantic meridional overturning circulation.
  The model includes a global ocean general circulation model and a statistical atmosphere model.
  The statistical atmosphere model is based on linear regressions of data from a fully coupled climate model on sea surface temperature both locally and hemispherically averaged, being the footprint of Atlantic meridional overturning variability. It provides dynamic boundary conditions to the ocean model for heat, freshwater and wind--stress.
  A basic but consistent representation of ocean--atmosphere feedbacks is captured in the hybrid coupled model and it is more than ten times faster than the fully coupled climate model.
  The hybrid coupled model reaches a steady state with a climate close to the one of the fully coupled climate model, and the two models also have a similar response (collapse) of the Atlantic meridional overturning circulation to a freshwater hosing applied in the northern North Atlantic.
\end{abstract}

\maketitle
\section{Introduction}
\label{sec:Introduction}

Since the pioneering work by~\cite{Stommel1961} on a conceptual model of the thermohaline circulation, the problem of the stability of the Atlantic Meridional Overturning Circulation (AMOC) has become one of the main issues in climate research. 
A collapse of the AMOC is often used to explain abrupt changes in past climate records. In recent years, a possible AMOC collapse in response to increased freshwater forcing in the northern North Atlantic, expected as a consequence of global warming, has been identified as a low probability but high risk future climate event \cite[]{Broecker1997,Clark2002,Alley2003}. 

An abrupt collapse of the AMOC, in response to a quasi--equilibrium increase in freshwater forcing in the North Atlantic, has been reported in different ocean and climate models of intermediate complexity (EMICs) \cite[]{Rahmstorf2005}.   
This  implies a non--linear response of the ocean to the freshwater forcing, with a sudden collapse of the overturning above a threshold value of the freshwater forcing.
The EMIC results are challenged by the model experiments of \cite{Yin2006}  and by IPCC--AR4 general circulation model (GCM) results, as analysed in \cite{Schmittner2005}. 
In the latter, it is found that the AMOC strength decreases approximately linearly in response to a  $CO_2$  increase according to the SRES--A1B scenario and there is no collapse.
It must be noted that the simulations to detect possible multiple equilibria regimes of the AMOC  in  these GCMs have not been done. 
The near--linear response to the gradual freshwater flux perturbation as found in  \cite{Schmittner2005}  does not rule out the possibility  of a sudden  collapse with a stronger freshwater flux. 

However, from the GCM results it has been suggested that the existence of a multiple equilibria regime is  an artifact of ocean--only  models, and  in particular of poor (or absent) representation of ocean--atmosphere interactions.
In an ocean--only model, the salt advection feedback is the central feedback affecting the stability of the AMOC\@.
When an atmosphere is coupled to the ocean model, other feedbacks, due to the ocean--atmosphere interaction,  become relevant.
The effect of these feedbacks may eventually overcome the effect of the salt--advection feedback,  and remove the multiple equilibria found  in ocean--only models and EMICS.

In some models, the response of the atmosphere to AMOC changes may indeed act to stabilise the present day AMOC \cite[]{Vellinga2002,Stouffer2006}.
In particular, the southward shift of the intertropical convergence zone would enhance the surface salinity of the Atlantic north of the equator, increasing the northward salinity transport by the northern hemispheric gyres~\cite[]{Krebs2007,Vellinga2002}.  The decrease in the atmospheric temperature of the Northern Hemisphere (NH), as a consequence of the AMOC collapse,  may also play a role~\cite[]{Stouffer2006}.
Lower atmospheric temperatures would determine stronger heat extraction  from the ocean and, consequently, higher densities of surface waters.
This effect may be more than compensated by the insulating effect of a NH ice cover extending more to the south~\cite[]{Vellinga2002}.
The potential impact of changes in the wind--stress, in particular zonal wind--stress, has recently been investigated in~\cite{Arzel2010}, but the magnitude of the changes induced by the wind--stress feedback remains unclear.

The question that must be answered is: ``Do the atmospheric feedbacks remove the multiple equilibria regime of AMOC,  as found in ocean--only models and EMICs?''
The first step to try to answer this question is, in our view,  to find a simple, but quantitative, description of these atmospheric feedbacks, extending that of box--model representations \cite[]{Nakamura1994}.
Only when a quantitative description of  the feedbacks is available, it is possible to assess the impact of the ocean--atmosphere interaction on the stability  properties of the AMOC. 
Studies to isolate the effect of the different feedbacks using a GCM  are computationally expansive.
Furthermore, the complexity of a full GCM can hinder the understanding of the relevant processes in the system.
For these reasons, simpler atmospheric models are needed to provide dynamic boundary conditions to full ocean GCMs.
Their design can benefit from the fact that the atmosphere, on the ocean time scales, can effectively be treated as a ``fast'' component that adjusts to the ocean anomalies.
These coupled models are often referred to as ``hybrid coupled models'' (HCMs). 

Since the main known atmosphere--ocean coupled mode of variability is the El Ni\~no Southern Oscillation (ENSO), HCMs have been developed mainly to study this phenomenon, focusing on the interaction between wind and sea surface temperature ($SST$) in the tropical oceans.
In this framework, the main atmosphere--ocean interaction  to include in the model is the change in the zonal winds over the equatorial Pacific in response to $SST$ anomalies \cite[]{Cane1986}.
\cite{Barnett1993} used a statistical model of the wind--stress based on an empirical orthogonal function decomposition of real data, coupled to a regional GCM of the equatorial Pacific.
They found good forecasting skill for ENSO variability prediction, and HCMs have been extensively used for ENSO forecasting since then \cite[]{Latif1998}.
Singular value decomposition of observational data has been used in \cite{Syu1995}, to implement an anomaly model of wind--stress for the equatorial Pacific. The HCM including this model has been used to investigate the role of ENSO--like feedbacks in seasonal variability.
In \cite{Burgers2003}, linear regressions on Ni\~no--3 and Ni\~no--4 indexes are used in combination with a red noise term to study the importance of local wind feedbacks in the Tropical Pacific.
Singular value decomposition in combination with a stochastic term has been used also in~\cite{Storch2005}.
In these studies, the wind--stress--$SST$ interaction is generally the main point of interest, but other feedbacks are active as well in the ocean--atmosphere system.
Changes in wind speed affect evaporation and, as a consequence, surface temperature \cite[]{Neelin1987}.
Also the freshwater flux is correlated to $SST$, through the triggering of convective events in the atmosphere \cite[]{Graham1987,Zhang1995}.

Our aim here is to develop a global HCM that includes all the main atmosphere--ocean feedbacks relevant for the stability of the AMOC, in an approach that focuses on the quasi--steady state behaviour rather than on variability.
As we want to follow an approach as general as possible, we regress all the surface fluxes pointwise on $SST$.
Since the $SST$ variability has a typical extent ranging from regional to basin scale, the atmosphere--ocean interaction is roughly captured by this local approach.
In the HCM,  two linear perturbation terms dependent on $SST$ are added  to the climatology of the forcing  fields of the ocean model.
A term depending on the local $SST$ anomaly represents the atmosphere--ocean  feedbacks that are acting in a statistical steady state.
The large--scale changes in the surface fluxes due  to the collapse of the AMOC can not be described by these local regressions alone, but  are included  through a second linear term that depends on the anomalous strength of the overturning circulation itself, measured through the NH annual average $SST$ anomaly.
Taken together, the local-- and large--scale terms give a simple representation of the atmospheric feedbacks which play a role in the stability of the AMOC.

As a demonstration of concept, our regressions are based on the output of an EMIC (described in section \ref{sec:Model}).
The linear atmospheric  feedback representations are presented in  section~\ref{sec:regressions} with results in section ~\ref{sec:Results}.
The performance of the HCM is compared to the one of the original EMIC in section~\ref{sec:test}.
With both local and large--scale  regression terms,  the HCM captures  the changes in atmospheric fluxes in response to AMOC changes.
The advantages of the HCM  over the EMIC  are that (i) a more than ten  fold decrease in computation time is achieved and (ii) it gives the possibility to selectively investigate the effect of different physical processes on the stability of the AMOC separately.

\section{The EMIC SPEEDO}
\label{sec:Model}

The HCM is constructed from data of the EMIC SPEEDO \cite[]{Severijns2009}, an intermediate complexity coupled atmosphere/land/ocean/sea--ice general circulation model. 
The choice for an EMIC is motivated by the fact that multi--thousand year runs are needed to construct the HCM, which is at the moment not feasible with a GCM.

The atmospheric component of SPEEDO is a modified version of Speedy~\cite[]{Molteni2003,Kucharski2003,Bracco2004,Hazeleger2005,Breugem2007}, an atmospheric GCM, having a horizontal spectral resolution of T30 with a horizontal Gaussian latitude--longitude grid (approximately $3^\circ$ resolution) and 8 vertical density levels.
Simple parameterisation are included for large--scale condensation, convection, radiation, clouds and vertical diffusion.
A simple land model is included, with three soil layers and up to two snow layers.
The hydrological cycle is represented with the collection of precipitation in the main river basins and outflow in the ocean at specific positions.
Freezing and melting of soil moisture is included.

The ocean model component of SPEEDO is the CLIO model~\cite[]{Goosse1999}. 
It has approximately a $3^\circ \times 3^\circ$ resolution in the horizontal, with 20 vertical layers ranging in resolution from $10\;m$ to $750\;m$ from the surface to the bottom.
The horizontal grid of the ocean model is curvilinear, and deviates from a latitude--longitude one in the north Atlantic and Arctic basins to avoid the singularity of the north pole.
A convective adjustment scheme, increasing vertical diffusivity when the water column is unstably stratified, is used in the model.
LIM sea--ice model is included in CLIO~\cite[]{Graham1987}.
A coupler provides the boundary conditions to the components, and performs the interpolations between the different ocean and atmosphere model grids in a conservative way.

Studies conducted both with an EMIC \cite[]{DeVries2005}  and with a fully implicit ocean model \cite[]{Huisman2010} showed the fundamental role of the salinity budget at the southern boundary of the Atlantic ocean in determining the response of the AMOC to freshwater anomalies \cite[]{Rahmstorf1996}.
The value of the net  freshwater transport  by the overturning circulation at $35^\circ\mathrm{S}$, shorthanded $M_{ov}$, is likely a control parameter that signals the coexistence of two stable equilibria of the AMOC.
If $M_{ov}$ is positive, the AMOC  is importing freshwater into the Atlantic basin and only the present--day ``ON'' state of the overturning is stable. If $M_{ov}$ is negative, freshwater is exported out of the basin by the AMOC, and a second stable ``OFF'' state of the AMOC exists, with reversed or no overturning in the Atlantic ocean. 

In the equilibrium solution of SPEEDO, the Atlantic basin integrated net evaporation is overestimated both with respect to most other models and to the few available observations~\cite[]{Rahmstorf1996}. 
Furthermore, the zonal gradient of salinity in the south Atlantic is reversed too, with a maximum on the eastern side. 
The high evaporation over the basin, combined with the low freshwater import by the gyre due to the reversed zonal salinity profile, force the overturning circulation to import freshwater ($M_{ov}=0.29 \; \mathrm{Sv}$) in order to close the budget.
For these reasons, a small freshwater flux correction is needed in the model for the purpose of our study, since we are interested in the feedbacks connected with a permanent collapse of the AMOC.
Following the example of \cite{DeVries2005}, a freshwater increase is applied over the eastern Atlantic, from the southern boundary to the latitude of  the Gibraltar strait, summing up to $0.2 \; \mathrm{Sv}$.
A dipole correction is applied over the southern gyre to reverse the zonal salinity profile, with a rate of $0.25 \; \mathrm{Sv}$\protect\footnote{The model used in~\cite{DeVries2005} shares the same ocean model component as SPEEDO, but uses ECBilt as the atmospheric model instead of Speedy. 
In their setup, the basin integrated net evaporation of the Atlantic ocean is underestimated, while the zonal salinity contrast in the southern Atlantic is overestimated. 
Therefore, their correction has a sign opposite to that here.}. 
All the corrections are performed as a virtual salt flux, keeping the global budget closed with an increased evaporation in the tropical Pacific and Indian oceans. 
As a consequence of these corrections, the net freshwater transport of the AMOC at the southern boundary of the Atlantic basin becomes negative ($M_{ov}=-0.069 \; \mathrm{Sv}$). 
As proposed in \cite{DeVries2005} and \cite{Huisman2010}, this situation may allow the coexistence of multiple equilibria of AMOC under the same boundary conditions.
Even if the data necessary for the definition of the HCM comes from 300 years of simulations alone, in the testing phase of different freshwater corrections applied to reach the regime where the MOC can permanently collapse, several tens of thousand years of integrations have been simulated by the EMIC (i.e., changing fresh-water correction and going to equilibrium, testing flux diagnostics, testing whether the collapse of the AMOC is permanent), motivating the use of a fast EMIC.

The surface boundary conditions for the ocean are computed from the atmospheric model as follows.
Since the atmospheric boundary layer is represented by only one model layer, near surface values of temperature ($T_{sa}$), wind ($\vec{U}_{sa}$, the bold font indicating a vector quantity) and specific humidity ($Q_{sa}$) are extrapolated from the values of the model lowest full layers. 
Furthermore, an effective wind velocity is defined to include the effect of unresolved wind variability as $\left | V_0 \right | = \left ( \vec{U}_{sa} \cdot \vec{U}_{sa} + V_{gust}^2\right)^\frac{1}{2}$, where $V_{gust}$ is a model parameter.
The ocean model provides through the coupler the values of $SST$, from which also the saturation specific humidity at the surface ($Q^{sat}_{sa}$) is computed through the Clausius--Clapeyron equation. 
With these quantities, the surface boundary conditions for the ocean are computed. 
The sensible ($\Phi_{SQ}$) and latent heat ($\Phi_{LQ}$) fluxes into the ocean are obtained from the bulk formulas:
\begin{equation}\label{eq:heat}
  \begin{split}
    \Phi_{SQ} &= \rho_{sa} c_p C_H \left | V_0 \right | \left(T_{sa} - SST \right ),\\
    \Phi_{LQ} &= \rho_{sa} L_H C_H \left | V_0 \right | min\left [\left(Q_{sa} - Q^{sat}_{sa} \right ),0\right],
  \end{split}
\end{equation}
where $\rho_{sa}$ is the surface air density, $c_p$ and $L_H$ are the specific heat of air and the latent heat of evaporation, respectively, and $C_H$ is a heat exchange coefficient, a model parameter depending on the stability properties of the boundary layer. 
The parameterisation of the radiative fluxes are more complex. 
For the short--wave ($\Phi_{SW}$) and long--wave components ($\Phi_{LW}$), two and four frequency bands are used, respectively. 
Transmittance is computed for each band separately, taking into account air density, water content and cloud cover.
The total non--solar heat flux ($\Phi_Q$) is just the sum of the different components:
\begin{equation}\label{eq:heatall}
  \Phi_{Q} = \Phi_{SQ} + \Phi_{LQ} + \Phi_{LW}.
\end{equation}
Separate parameterisation are used for precipitation due to convection ($\Phi_{Pcv}$) and to large--scale condensation ($\Phi_{Pls}$).
River runoff ($\Phi_R$) is provided by the land model.
The net evaporation ($\Phi_E$) can then be computed as:
\begin{equation}\label{eq:net_evap}
  \Phi_{E} = \Phi_{LQ}/L_H - \Phi_{Pls} - \Phi_{Pcv} - \Phi_{R}.
\end{equation}
The wind--stress vector is computed as:
\begin{equation}\label{eq:wind_stress}
  \vec{\Phi_U} = \rho_{sa} C_D \left | V_0 \right | \vec{U}_{sa},
\end{equation}
where $C_D$ is a drag coefficient.

\section{Linear regressions}
\label{sec:regressions}

Our aim is to capture the changes in the atmospheric forcing connected with the changes in the ocean state, that is the atmospheric response to a collapse of the AMOC. 
As motivated in the introduction, we assume that these atmospheric feedbacks can be expressed as functions of $SST$ alone.  
First, the feedbacks that keep the system in a statistical equilibrium state are always present, and are expressed in our case as a function of local $SST$.
They are extracted from a 200 years long statistical steady state run (CLIM) of SPEEDO.
The departure from the steady state arises during an externally forced AMOC collapse, in association with the large--scale $SST$ footprint of a AMOC decline. 
The feedbacks involved in the collapse are different from the ones acting at the steady state.
To study the large--scale feedbacks, a 4000 year experiment was performed, starting from CLIM, with an additional $0.4 \;\mathrm{Sv}$ freshwater flux centred around southern Greenland during the first 1000 years; this run is referred to as PULSE.
In the first hundred years of the experiment, the AMOC collapses and a shallow reverse overturning cell is established in the Atlantic basin. 
Since in this paper the focus is only on the impact of a complete and steady collapse of the AMOC, we only show the results using the large freshwater anomaly mentioned, that guarantees that the AMOC is brought to a steady reversed state.

\begin{figure}[tbp]
  \begin{center}
    \includegraphics[width=0.8\columnwidth]{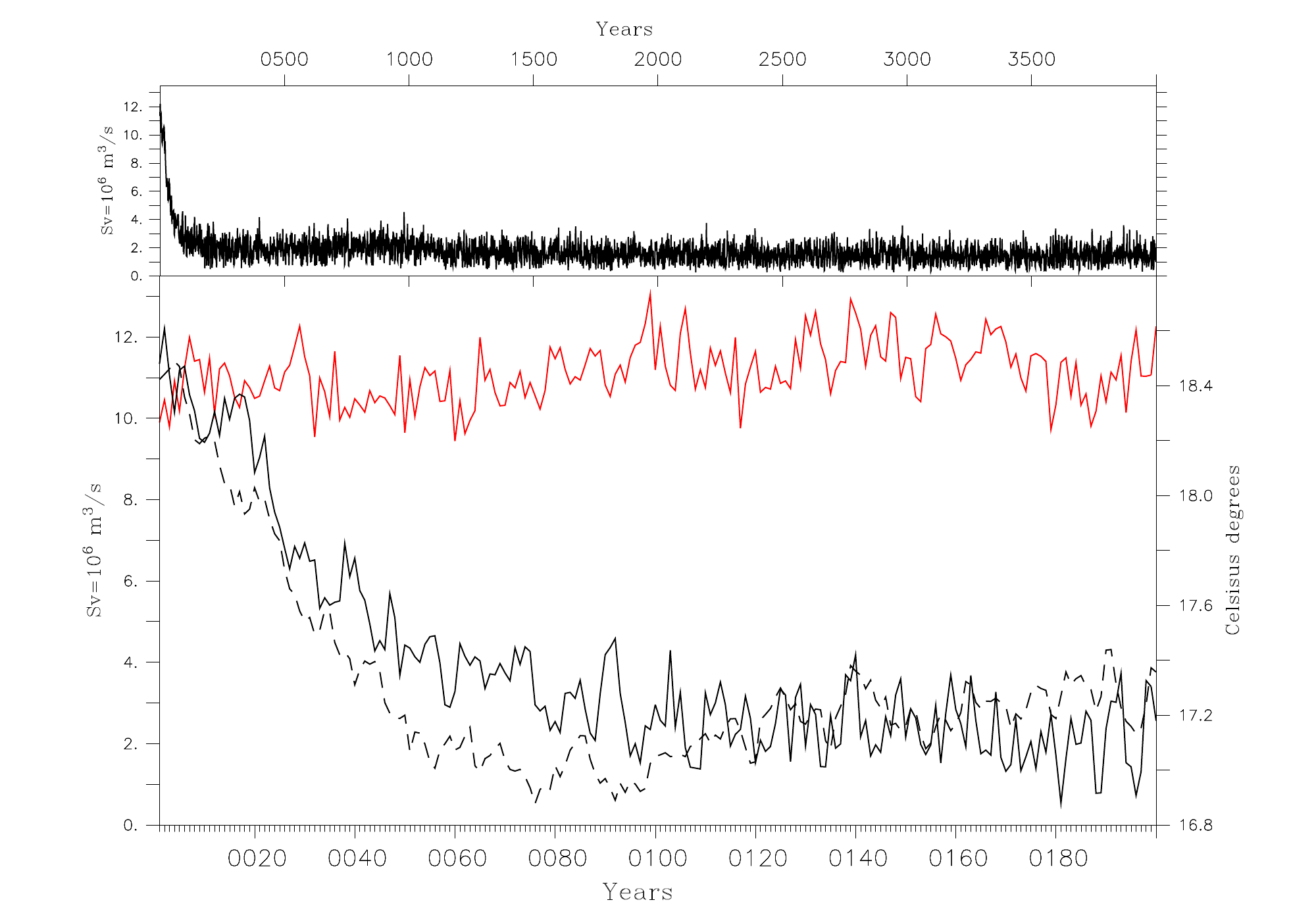}
  \end{center}
  \caption{
  Top panel: Maximum AMOC for the entire PULSE experiment, in $\mathrm{Sv}$.
  Bottom panel: Maximum AMOC for CLIM (red full line) and first 200 years of PULSE (black full line) in $\mathrm{Sv}$ (left $y$--axis). NH average $SST$ in $^\circ \mathrm{C}$ (right $y$--axis) for PULSE (dashed black line).
  }
  \label{fig:streams}
\end{figure}

The maximum of the meridional overturning streamfunction during the first two hundred years of both PULSE and CLIM runs are shown in figure~\ref{fig:streams} (bottom panel).
After the first 1000 years of the experiment, the additional freshwater pulse is released and the model tends to an equilibrium state with no sign of recovery of deep water formation in the northern north Atlantic after 3000 years (top panel of figure~\ref{fig:streams}).
Taken together, the feedbacks extracted from CLIM and PULSE runs provide the representation of the changes of the atmospheric fluxes during a collapse of the AMOC.

To provide the simplest description of the changes taking place at the ocean--atmosphere interface, the first order approximation is the addition of a linear perturbation term to the climatology of surface atmosphere--ocean fluxes. 
In particular, we consider a linear regression on $SST$\@. 
This approach is clearly limited, but it is an approximation that gives a consistent representation of the large--scale feedbacks.
The results can be successfully used as boundary conditions for the ocean--only model, as will be shown below.

To force the ocean model, we need five surface fluxes: non--solar heat flux (that includes long--wave radiation, latent and sensible heat fluxes), short--wave radiative heating, net evaporation, zonal and meridional wind--stresses. 
The incoming short--wave radiation is not regressed, and only its average seasonal cycle is retained, since its response to SST is completely mediated through a cloud cover response that is not well represented in the Speedy model~\cite[]{Severijns2009}.

Two linear models are used for regressing data from CLIM and PULSE.
The CLIM data is fitted with:
\begin{equation}\label{eq:reg} 
  \phi(i,j) -  \overline{\phi(i,j)} = p_1(i,j)\cdot \left(SST(i,j)-\overline{SST(i,j)}\right),
\end{equation}
where $\phi \in \left \{\Phi_Q,\Phi_E,\vec{\Phi_U} \right\}$ is a particular surface flux field to be regressed, $p_1$ is the model parameter field to be fitted, $i$ ($j$) is the grid index in the east--west (north--south) direction and the overbar indicates a time average. 
Monthly data is used in the fit of CLIM data to represent the seasonal cycle. 
Note that this formulation is a \emph{local} regression, by which we mean a regression between quantities that belong to the same grid cell of the model.

The natural variability signal caught by regressions from equation~(\ref{eq:reg}) is removed from PULSE data. 
Only the first 100 years of PULSE are used, since we are interested in the response that can approximately be considered linear. 
The residual signal $\phi_r(i,j)$ can then be regressed with a second linear model:
\begin{equation}\label{eq:reg_LS}
 \phi_r(i,j) = p_2(i,j) \cdot \left( \left < SST\right >_{NH} - \overline{\left < SST \right >_{NH}}\right ),
\end{equation}
where the symbol $\left < \;\;\; \right>_{NH}$ denotes the average over the NH.
In this case the regressor is, for all grid cells, the yearly average SST in the NH, a good indicator of the state of the AMOC~\cite[]{Stouffer2006}, as figure~\ref{fig:streams} suggests (bottom panel, dashed line). 
Yearly mean data is used for the fit of PULSE.
It must be stressed that the last term of equation~(\ref{eq:reg_LS}) is the average NH $SST$ for the CLIM run, since we are interested in the deviation from the equilibrium state.
Consequently, the intercept is set to zero, since the terms involving $p_2$ need not to have an effect when the climate is in a neighbourhood of CLIM.

All the regressions are computed with the \emph{lm} (linear model) function provided in the R statistical software, version 2.8.0 \cite[]{Team2009}. 
The regressions are computed through a least square technique, and we require a statistical significance higher than the 95 percentile, discarding all the fits with a \emph{p--value} (provided by \emph{lm} itself) higher than $0.05$. 
This equals to discarding a fit if the probability of having the same result using random data is higher than 5\%.
When this occurs the fit is considered unsuccessful, and only the climatological value of CLIM ($\overline{\phi(i,j)}$ in equation~(\ref{eq:reg})) is kept and both $p_1(i,j)$ and $p_2(i,j)$ are set to zero.
The output of the fitting procedure shows very weak sensitivity to the chosen significance level.

The same regression procedure was applied also to the output of the uncorrected original SPEEDO model.
The results obtained from the two models, with or without freshwater flux corrections, are consistent on both qualitative and quantitative grounds.
A partial exception is the southern ocean and the Labrador sea, where the strength of the feedbacks is different.
An analysis of these differences is beyond the scope of the present study, but may be associated with changes in sea--ice cover in the two models.

We now give the formulation of the boundary conditions for the ocean--only model to be forced by our ``climatology with feedbacks''.
The surface heat flux into the ocean is computed as a combination of the regressions and a restoring term to the climatology:
\begin{equation}\label{eq:heat_forcing} 
  \begin{split}
    \Phi_{Q}(i,j) =&  \overline{\Phi_{Q}(i,j)} + p_1^{\Phi_{Q}}(i,j)\cdot \left(SST(i,j)-\overline{SST(i,j)}\right) \\
                +& p_2^{\Phi_{Q}}(i,j) \cdot \left( \left < SST\right >_{NH} - \overline{\left < SST \right >_{NH}}\right ) \\
		+& \overline{\Phi_{SW}(i,j)}\\
		+&  \frac{\rho_{sa} c_p \left | \overline{V_0(i,j)} \right |}{\tau} \cdot \left ( \overline{SST}(i,j)-SST(i,j) \right ),
  \end{split}
\end{equation}
where $p_1^{\Phi_Q}$ and $p_2^{\Phi_Q}$ are the local and large--scale regression parameters for the heat flux, $\rho_{sa}$ and $\overline{V_0(i,j)}$ are fixed climatological values and the relaxation time $\tau$ is chosen to be 55 days for the ocean, consistently with the bulk formula of the coupled model of equation~(\ref{eq:heat}).

The net evaporation flux is computed in three steps. 
First, the deviations from the climatological values, $\delta \Phi_E$, are computed at each grid cell: 
\begin{equation}\label{eq:evap_forcing}
  \begin{split}
    \delta \Phi_{E}(i,j) =& p_1^{\Phi_{E}}(i,j)\cdot \left(SST(i,j)-\overline{SST(i,j)}\right) \\
                +& p_2^{\Phi_{E}}(i,j) \cdot \left( \left < SST\right >_{NH} - \overline{\left < SST \right >_{NH}}\right ),
  \end{split}
\end{equation}
where $p_1^{\Phi_{E}}$ and $p_2^{\Phi_{E}}$ are the regression parameters for the net evaporation flux.
Then, the global integral of the deviations, $\Delta \Phi_E$, is computed on the model grid and the budget imbalance is set to zero. 
The total freshwater flux reads then:
\begin{equation}\label{eq:evap_total}
    \Phi_{E}(i,j) = \overline{\Phi_{E}(i,j)} + \delta \Phi_E(i,j) - \Delta \Phi_E/\Sigma,
\end{equation}
where $\Sigma$ is the ocean surface area.

For the wind--stress vector, only the output of the regressions is used:
\begin{equation}\label{eq:wind_forcing} 
  \begin{split}
    \vec{\Phi_U}(i,j) &=  \overline{\vec{\Phi_U}(i,j)} + \vec{p}_1^{\vec{\Phi_U}}(i,j)\cdot \left(SST(i,j)-\overline{SST(i,j)}\right) \\
    &+ \vec{p}_2^{\vec{\Phi_U}}(i,j) \cdot \left( \left < SST\right >_{NH} - \overline{\left < SST \right >_{NH}}\right ),
  \end{split}
\end{equation}
where $\vec{p}_1^{\vec{\Phi_U}}(i,j)$ and $\vec{p}_2^{\vec{\Phi_U}}(i,j)$ are the vectors of the regression parameters for local and large--scale regressions respectively, for the two components of the wind--stress.
Over sea--ice, a fixed climatology of air--ice fluxes is used. 
When sea--ice is present, weighting is applied by the model to the surface fluxes multiplying by the fractional ocean area $(1-\varepsilon(i,j))$, where $\varepsilon(i,j)$ is the fractional sea--ice cover of the cell.

The technique described returns the rate of change of the field with $SST$ or $\left < SST\right >_{NH}$ only in those areas where a linear regression is statistically significant.
Furthermore, setting the regression parameters to zero still leaves a constant climatology that can be used as boundary condition for the ocean model.
We thus have the complete control over which feedbacks are acting at the ocean--atmosphere interface, and we can selectively investigate their individual or collective effect.

\section{Results}
\label{sec:Results}

\subsection{Local regressions}
\label{sec:localreg}

\begin{figure}[tbp]
  \begin{center}
    \includegraphics[height=0.8\textheight]{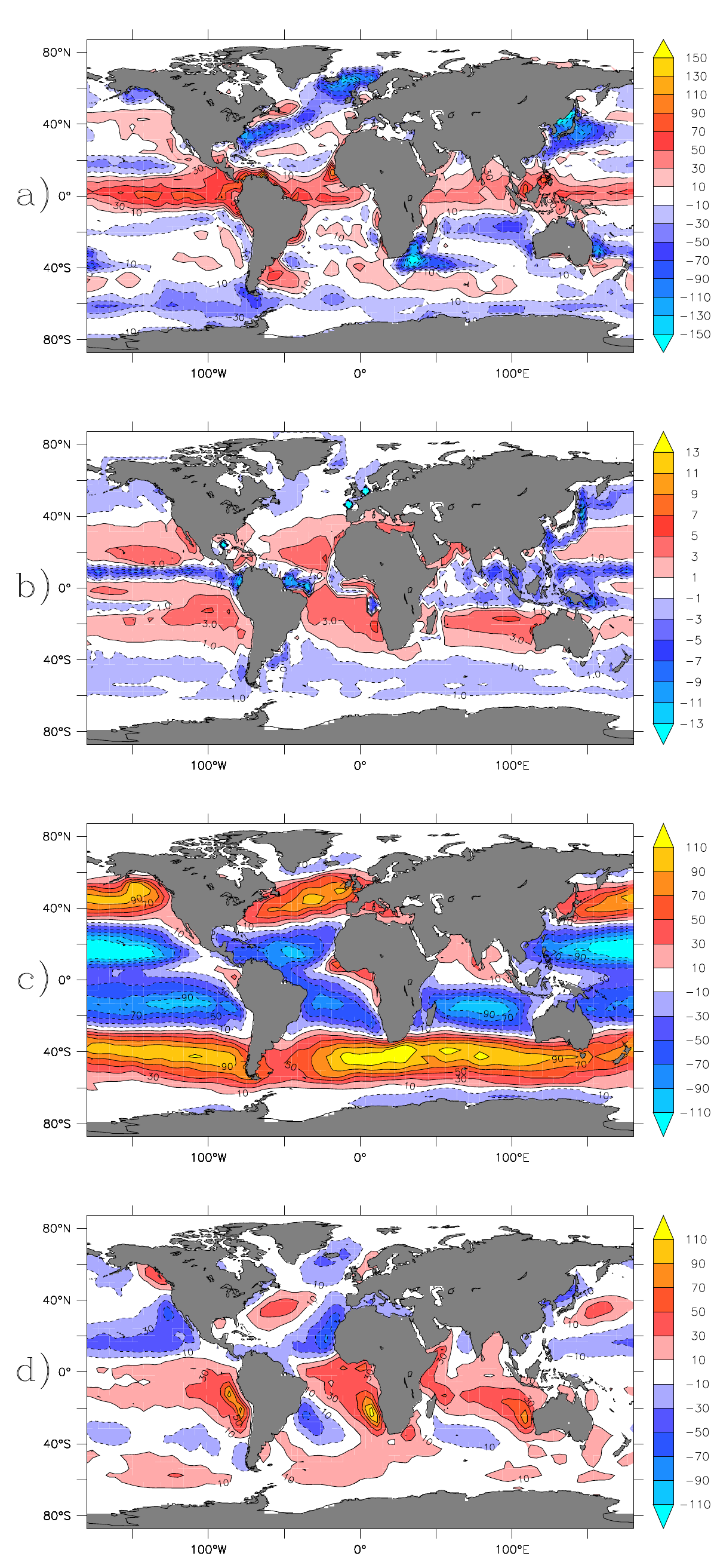}
  \end{center}
  \caption{Average value of the regressed fields from CLIM data ($\overline{\phi(i,j)}$ in equation~(\ref{eq:reg})), weighted by the fractional ocean area $\overline{(1-\varepsilon(i,j))}$. a) Total heat flux in $W/m^2$, positive downwards. b) Net evaporation in $mm/day$. c) and d) are the zonal and meridional components of wind--stress respectively, in $10^{-3}\cdot N/m^2$.}
  \label{fig:Climatology}
\end{figure}

\begin{figure}[tbp]
  \begin{center}
    \includegraphics[height=0.8\textheight]{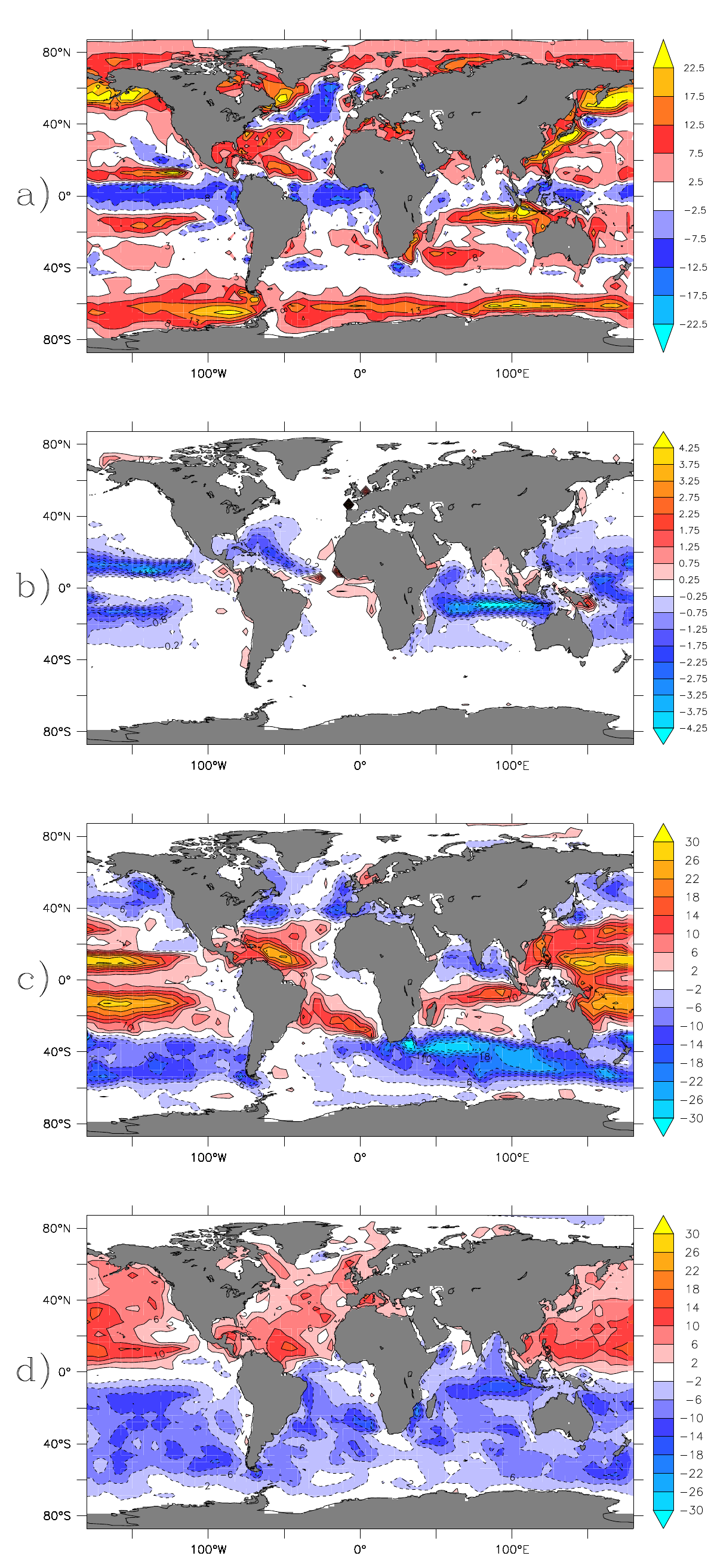}
  \end{center}
  \caption{As in figure~\ref{fig:Climatology}, but for the local regression parameter $p_1$. The units are the same of figure~\ref{fig:Climatology}, divided by $^\circ \mathrm{C}$. In panel a, only the non--solar heat flux is considered.}
  \label{fig:p1_reg}
\end{figure}

The fitting procedure for CLIM data is generally successful and the results of the regressions on CLIM data are reported in figures~\ref{fig:Climatology} and~\ref{fig:p1_reg}. 

In figure~\ref{fig:Climatology}, the average value of the regressed fields is reported ($\overline{\phi(i,j)}$ in equation~(\ref{eq:reg})). The total heat flux (including short--wave radiation) is shown in figure~\ref{fig:Climatology}. The net evaporation includes the river runoff. 
The values of the regression parameter $p_1$ are shown in figure~\ref{fig:p1_reg} for all the regressed fields. 
In both figures~\ref{fig:Climatology} and~\ref{fig:p1_reg} the values are weighted by the fractional free ocean surface of the cell to compensate for the effects of average sea--ice cover.
The effect of changes in sea--ice cover are not included into the regressions, as the effect of sea--ice is taken into account by CLIO model.
As discussed below, the changes in sea--ice can strongly modify the feedbacks (compare figures~\ref{fig:p1_reg} and~\ref{fig:reg_effective}).

For all the regressed fields, the contribution to the fluxes of the local regression terms can be important compared to the average value, in particular at the western boundaries and outside the equatorial and polar regions.
This is clear when we consider the $SST$ variability on a daily basis; the root of the variance is well above $1^\circ \mathrm{C} $ everywhere in the subtropical and subpolar ocean, with peak values of about $7 ^\circ \mathrm{C}$ close to the NH western boundaries (not shown).

The linear regressions only capture part of the natural variability of CLIM fluxes, but the error is generally lower than $10\%$ of the original field over a major part of the ocean (not shown). 

Apart from the standard damping on $SST$ that also operates in ocean--only models driven by a prescribed atmosphere, the atmospheric control over the atmosphere--ocean heat flux counteracts this damping in many regions, in particular in the tropics and at high latitudes (positive values in figure~\ref{fig:p1_reg} a). 
This means that the linear feedback for the heat flux is not damping the $SST$ anomalies.
Relevant exceptions are the equatorial ocean, the central north Atlantic, the northern portion of the Southern Ocean and other smaller areas.
It should be noted that in the polar areas, the sea--ice cover determines the effective feedback in the heat flux, and often changes the sign of the feedback.
The exact mechanism of this feedback is discussed in more detail in section~\ref{sec:largereg}.

To investigate the origin of the pattern of the local heat feedback outside the polar regions, the same regression procedure was applied to each component of the heat flux separately, namely sensible and latent heat fluxes and long--wave radiation (not shown).
The change in the latent heat release is the most important component of the heat flux change.
The feedback of sensible heat flux is slightly weaker in magnitude, and is positive with the only relevant exceptions of the North Atlantic and the equatorial ocean.
The long--wave radiation feedback follows the same pattern, and is the weakest term.
As first noted in~\cite{Frankignoul1985}, the sign of the heat flux feedback from equation~(\ref{eq:heat}) depends to first order only on the relative change of $T_{sa}$ and $SST$, if the wind is assumed constant.
A positive feedback is possible only if the change in $T_{sa}$ is larger than the one in $SST$.
This is almost always true in our model in the areas where the heat feedback is positive, as we find when $T_{sa}$ is regressed on $SST$ (not shown).

A plausible explanation of this positive heat feedback, at least at low and mid latitudes, is given by the convection--evaporation feedback mechanism as proposed by~\cite{Zhang1995}.
There is a strong resemblance between the patterns of increased convective precipitation and those of weaker latent heat loss at higher $SST$.
This suggests that, in the tropical and subtropical areas where a positive heat flux feedback is observed, a positive $SST$ anomaly is associated with anomalous convergence of wet air that both contributes to the reduction of evaporation\protect{\footnote{The reduction of evaporation is mainly due to weaker winds.}} and enhances precipitation if convection is triggered.
Regression of surface pressure on $SST$ also supports this hypothesis, since higher $SST$s correlate with lower surface pressure in the tropical and subtropical areas.
Regarding net evaporation (figure~\ref{fig:p1_reg} b), a weak increase is observed at higher $SST$ over most of the ocean.
On the contrary, in most of the tropical areas the increase in convective events leading to increased precipitation dominates the freshwater feedback (basically, the blue areas of figure~\ref{fig:p1_reg} b), as discussed above.

In the case of wind--stress, a decreased magnitude is observed in connection with higher $SST$ (compare figure~\ref{fig:p1_reg} c and d with the mean fields of figure~\ref{fig:Climatology}).
The term $\left | V_0 \right |$ of equation~(\ref{eq:heat}) is regressed on the local $SST$, confirming that over most of the ocean at low and mid latitudes lower than average winds are observed in association with higher than average $SST$s (not shown), implying lower heat transfer through the interface.
The correlation decreases moving poleward and the mechanism involved is basically the wind--evaporation feedback~\cite[]{Neelin1987}, that connects higher evaporation (lower $SST$) with stronger winds.
The fact that we do not observe stronger winds where an increase of convective precipitation is found is not surprising, since the parameterisation of convection does not affect the horizontal wind field~\cite[]{Molteni2003}.
A positive correlation between wind speed and $SST$ is observed only in the western part of the subtropical gyre of the Southern Hemisphere (SH) of the Atlantic ocean, south of Greenland and in the Labrador sea, in the northeastern part of the subpolar gyre of Pacific ocean, and in some other smaller regions.
Even though the negative wind feedback is thought to be dominant, some evidence for a positive feedback has been found for the Kuroshio extension area, in the northeastern Pacific~\cite[]{Wallace1990,Nonaka2003}.
The best known wind--$SST$ feedback mechanism where the wind response to SST anomalies is central is the Bjerknes' feedback in the equatorial Pacific areas, in connection with the ENSO~\cite[]{Cane1986}.
The fundamental coupled variability of the equatorial ocean--atmosphere system is that of a decrease of the western Pacific trade winds in response to a positive anomaly of $SST$ in the eastern equatorial Pacific.
Even though the model has too low resolution to exhibit a realistic ENSO~\cite[]{Severijns2009}, a weakening of the trade winds in the western and central equatorial ocean is captured by the linear regressions (figure~\ref{fig:p1_reg} c) and is consistent with the anomaly patterns connected with ENSO~\cite[]{Deser1990}.
The stronger convective precipitation detected in the western Pacific at higher $SST$s may be a sign of anomalous convergence of the low level atmospheric circulation, again in agreement with what shown by~\cite{Deser1990}.
The origin of the dipole structure of the meridional wind feedback between NH and SH (figure~\ref{fig:p1_reg} d) is basically a reflection of the weaker dominant winds at higher $SST$.

\subsection{Large--scale regressions}
\label{sec:largereg}

Moving to the results of \emph{large--scale} regressions, it must be kept in mind in the interpretation of the results that the fit is performed only on the residuals of \emph{local} regressions, not on the full data of PULSE and that the fit is performed on a decreasing quantity, the NH average $SST$.

\begin{figure}[tbp]
  \begin{center}
    \includegraphics[height=0.8\textheight]{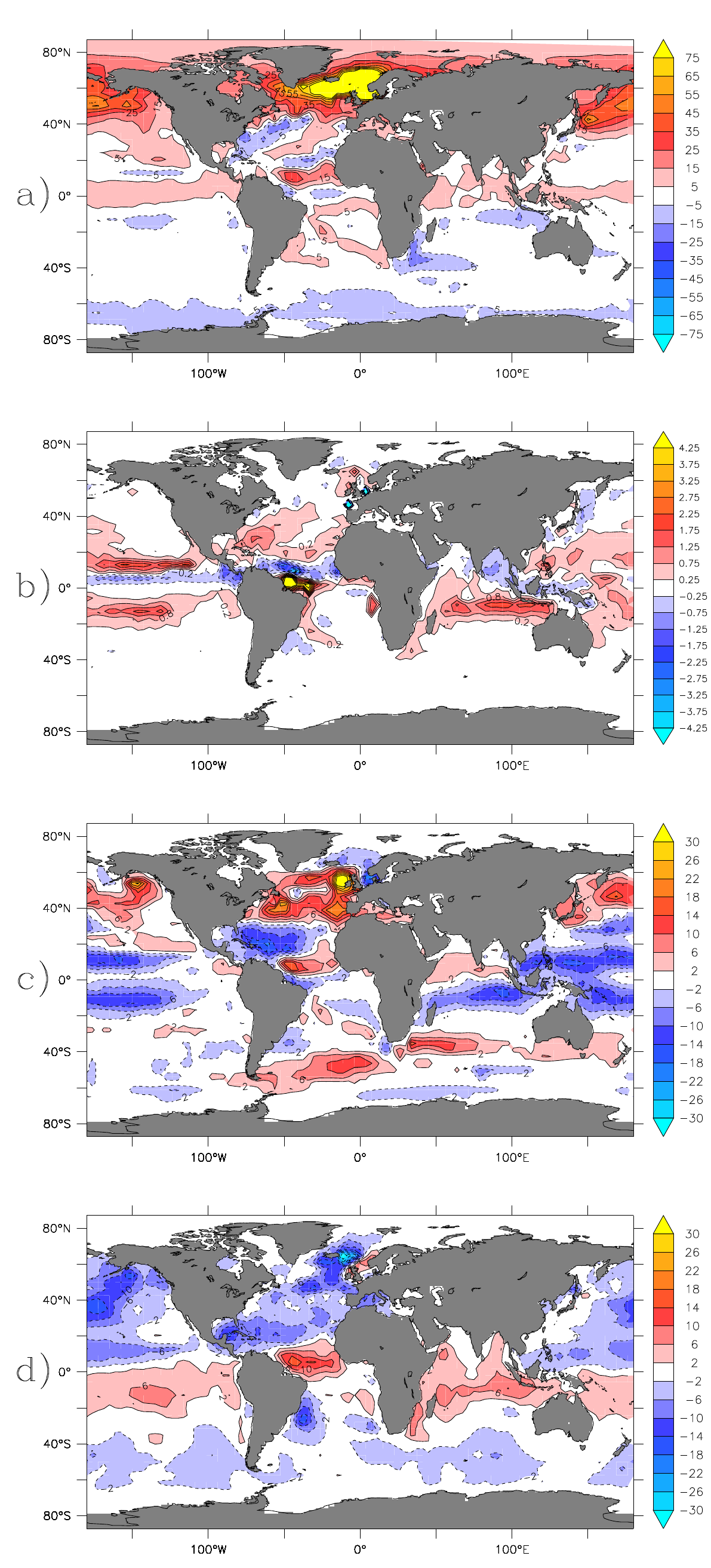}
  \end{center}
  \caption{As in figure~\ref{fig:Climatology}, but for the large--scale regression parameter $p_2$. The units are the same of figure~\ref{fig:Climatology}, divided by $^\circ \mathrm{C}$ In panel a, only the non--solar heat flux is considered. In panel b, the signal of the freshwater pulse has been removed from the source data.}
  \label{fig:p2_reg}
\end{figure}

\begin{figure}[tbp]
  \begin{center}

    \includegraphics[width=0.8\columnwidth]{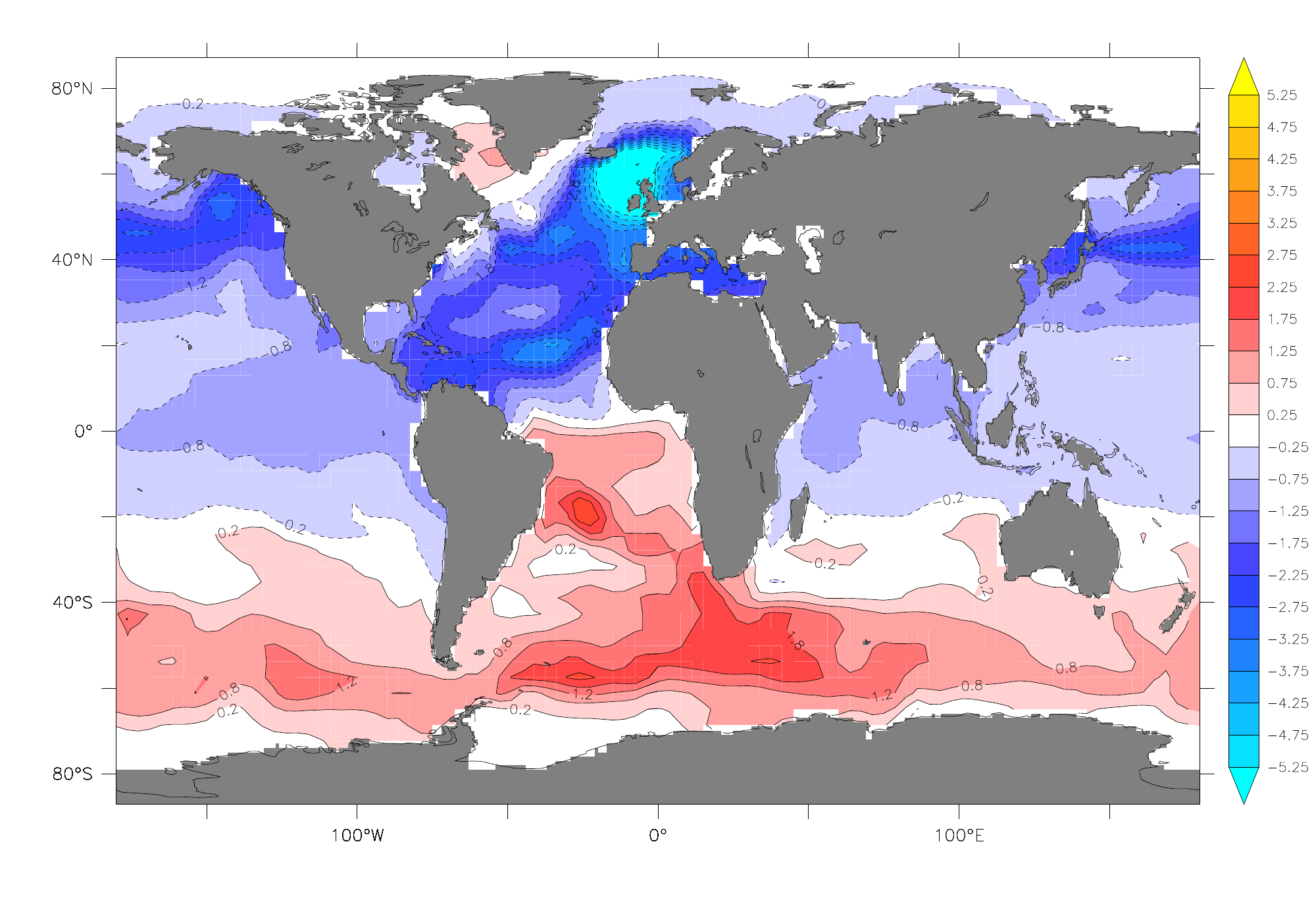}
  \end{center}
  \caption{
  Difference in $SST$ ($^\circ \mathrm{C}$) between the years 91--100 of PULSE experiment and the mean state of CLIM.}
  \label{fig:sst_variance}
\end{figure}

\begin{figure}[tbp]
  \begin{center}
    \includegraphics[height=0.6\textheight]{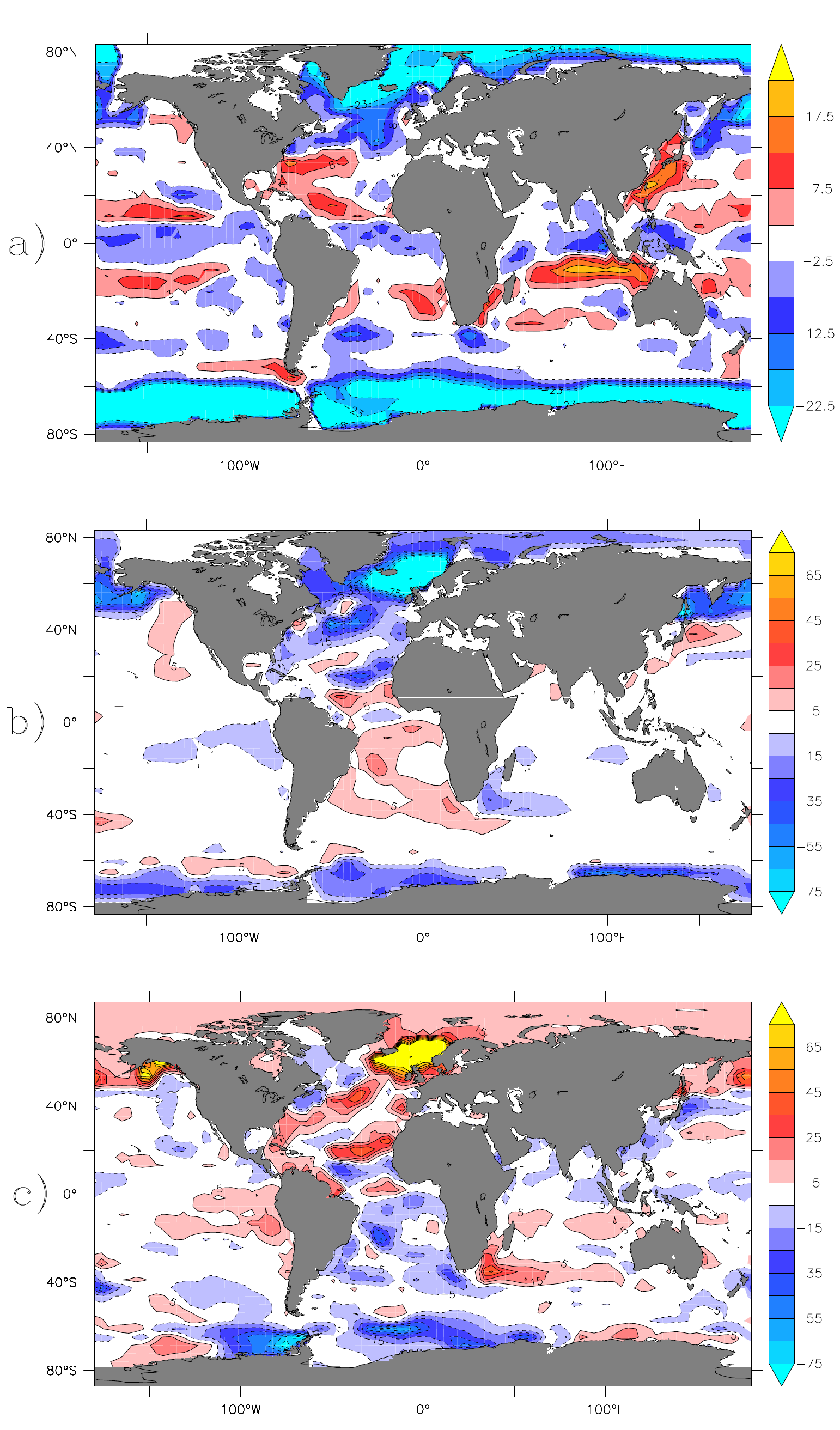}
  \end{center}
  \caption{Effective feedbacks for heat flux (short wave radiation excluded), when changes in sea--ice are considered (see text). Effective regression parameters for the heat flux, computed including the effect of changes in sea--ice ($p_1$ in panel a and $p_2$ in panel b). Also the change in the heat flux, as directly diagnosed from the coupled model, is shown in panel c, computed as the difference in ice--weighted heat flux from years 91-100 and 1-10 of PULSE.
Note that a different color scale is used in the top panel.
}
  \label{fig:reg_effective}
\end{figure}

The collapse of the AMOC causes a decrease in the NH average $SST$ of about $1.2 ^\circ \mathrm{C}$.
A weaker change of opposite sign is observed over the Southern Ocean (approximately $0.4 ^\circ\mathrm{C}$).
This NH--SH temperature dipole is a robust feature of different models, and is connected with lower northward heat transport in the Atlantic ocean, as already found in \cite{Stouffer2006}. 
The changes in the heat flux are mainly captured by the large--scale regression parameter alone.
This can be evinced comparing the large--scale heat flux parameter and the diagnosed changes in the flux from the coupled model, and is connected with the larger magnitude of the large--scale parameter.
The main response of the heat flux after the overturning collapse, not considering changes in the sea--ice cover (figure~\ref{fig:p2_reg} a), would be that of an increased heat extraction from the ocean in the NH ($9.9 \; W/(m^2 \;^\circ \mathrm{C})$ on average). 
When the effect of a changing sea--ice cover is included in the computation of the heat feedback (figure~\ref{fig:reg_effective}~b), its sign changes in the high latitudes of the NH ($-9.6 \; W/(m^2 \;^\circ \mathrm{C})$ on average in the NH), which means that heat released to the atmosphere decreases.
This result is in contrast with what the regression parameter $p_2$ suggests, but consistent with the sign of the effective regression parameter.
The difference is explained below.
The net heat flux, weighted by the ice--free area $(1-\varepsilon)$, can be written as:
\begin{equation}
  \phi_Q=(1-\varepsilon)(\overline{\phi_{Q}} + \partial\phi_Q/\partial SST + \partial\phi_Q/\partial \left<SST\right>_{NH}).
\end{equation}
$p_2^{\phi_Q}$ is simply $\partial\phi_Q/\partial \left<SST\right>_{NH}$ while the effective parameter is:
\begin{equation}
  \begin{split}
    p_{2,eff}^{\phi_Q} &=\partial(\phi_Q\cdot(1-\varepsilon))/\partial \left<SST\right>_{NH}\\
              &=(1-\varepsilon)\partial\phi_Q/\partial \left<SST\right>_{NH} - \phi_Q \partial\varepsilon/\partial \left<SST\right>_{NH}\\
              &=(1-\varepsilon)p_2^{\phi_Q} - \phi_Q \partial\varepsilon/\partial \left<SST\right>_{NH}.
  \end{split}
  \label{eq:effective}
\end{equation}
The second term on the right hand side of equation~\ref{eq:effective} describes the changes in sea--ice cover in response to $SST$ changes. This term is larger than the first term over most of the Northern North Atlantic.
Sea--ice cover changes determine the sign change in the large--scale heat feedback term.
A similar reasoning holds for the local feedback.
In general, the NH--SH heat flux dipole seen in figure~\ref{fig:p2_reg} a is driven by the decrease of NH near--surface temperature, that follows a pattern similar to that of $SST$ (figure~\ref{fig:sst_variance}), but with stronger sensitivity to AMOC changes everywhere except for the southern mid latitudes.
This amplification of the $SST$ signal, in particular in the atmosphere of the high latitudes of NH, is a consequence of the appearance of sea--ice during winter.
Without sea--ice changes, these differential variations in $SST$ and atmospheric temperature would tend to produce an increased upward heat flux in the NH (figure~\ref{fig:p2_reg} a).
This increased heat loss is more than counteracted by the decrease in open ocean area; the increased ice cover effectively drives the cooling of amospheric temperatures above the North Atlantic.
This can be seen from the changes in the heat flux diagnosed from the coupled model including the insulating effect of sea--ice (figure~\ref{fig:reg_effective}~c) and this is confirmed by the \emph{large--scale} regression parameter computed including the effect of sea--ice (figure~\ref{fig:reg_effective}~b).
This ``effective'' regression parameter is the result of the same fitting procedure, applied in this case to the surface heat flux weighted by the actual sea--ice cover and not to the complete heat flux.
The results for the local (large--scale) regression are those shown in figure~\ref{fig:reg_effective} a (b).
As a consequence, this regression parameter gives a better representation of the feedbacks that the ocean effectively senses (including the effect of sea--ice).
Note that the HCM only uses $p_1$ and $p_2$, and not the effective response coefficients.
The changes in sea--ice cover result from explicitly resolved ice dynamics and thermodynamics.

At low and mid latitudes in the NH the changes are due to reduced evaporation in response to lower $SST$ and, at low latitudes, to lower wind speed.
The changes in the surface long--wave radiation budget are smaller in magnitude, and amount to an increased net emission of long--wave radiation almost everywhere in the NH except from the GIN seas.
This effect has been observed in other model experiments and is connected with the reduced downward long--wave radiation flux over compensating the decreased black body emission at lower $SST$s~\cite[]{Laurian2009}.
The decrease in the downward long--wave flux is an effect of a drier atmosphere, and partly balances the reduced latent heat flux.
These changes in heat flux amount to a positive feedback on an AMOC anomaly when the effect of sea--ice is included, favouring a decrease of the surface density in the deep water formation areas of the North Atlantic in connection with weaker overturning circulation. 

The patterns of the net evaporation change (figure~\ref{fig:p2_reg} b) are consistent with the findings of~\cite{Stouffer2006} (their figure 9 e, with opposite sign).
The AMOC collapse causes a reduction of net evaporation over the tropical and subtropical NH and over the tropical SH, due to lower $SST$s (figure~\ref{fig:sst_variance}).
In the few areas where an increase in evaporation is observed (basically the north equatorial oceans), this is due to stronger winds. 
At low latitudes, a significant change of the precipitation patterns also plays a role, with a dipole pattern centred around the equator, and positive to the south.
This southward shift of the intertropical convergence zone (ITCZ) produces the strongest precipitation increase over the Amazon river basin.
This response of the Hadley cell is connected with the southward shift of the latitude of maximum heating, and has been observed consistently in different climate models~\cite[]{Stouffer2006,Krebs2007,Laurian2009} and in an idealised framework too~\cite[]{Drijfhout2010}.
A similar, though weaker, pattern of precipitation change is observed in the Pacific and Indian oceans.
The increased precipitation over the entire southern Atlantic more than compensates for the increased evaporation due to higher $SST$ in this part of the basin.
A slow down of the hydrological cycle over Europe is detected as two negative peaks off the coast of France and in the North sea.
On a global scale, the regressions of PULSE residuals determine an evaporation increase of $0.13 \; mm/(day \; ^\circ \mathrm{C})$. 
Therefore, our linear approach is not conserving the ocean water mass and needs a budget closure correction when used as boundary condition for the ocean, as implemented in equation~(\ref{eq:evap_total}).

In the case of wind--stress, the response of the atmosphere is somewhat less straightforward to understand, and it deserves a longer discussion.
For what concerns the meridional wind--stress, the changes in the low and mid latitudes are driven by the response of the zonally averaged temperature profile to the AMOC collapse.
The equator to pole temperature difference increases by approximately $4^\circ \mathrm{C}$ in the NH.
In the SH, the opposite is true, with a smaller change.
These changes are clearly mirrored in the zonally averaged wind--stress.
Stronger southward wind blows on the ocean with a collapsed AMOC in the NH up to $50^\circ \mathrm{N}$. 
The situation is similar in the SH, but with a weaker circulation down to $40^\circ \mathrm{S}$, following the opposite change in the zonally averaged temperature.
The zonal winds over the Southern Ocean are also reduced.
A more peculiar feature is observed in the north Atlantic.
A pressure anomaly dipole between Greenland and northeastern Atlantic develops, with positive sign to the east, in connection with the differential cooling between these two regions (stronger cooling over eastern Atlantic).
This in turn determines an anomalous anticyclonic circulation centred north of Scotland, with impacts on both the meridional and zonal wind--stress.
Referring to our regressions, the changes due to the AMOC collapse in the tropical regions are already caught by the local regression parameter ($p_1$, figure~\ref{fig:p1_reg} c and d).
This can be understood considering that the change in $SST$ due to the AMOC collapse (figure~\ref{fig:sst_variance}) is a dipole centred at the latitude of the southern tropic (at the equator in the Atlantic ocean) and positive to the south of it, with an amplitude of a few degrees.
In fact, the changes due to the overturning collapse are overestimated by the \emph{local} regressions, and $p_2$ (figure~\ref{fig:p2_reg} c and d) amounts to a correction opposite to $p_1$.
The positive values of $p_2$ for meridional wind--stress in the intertropical regions (figure~\ref{fig:p2_reg} d) signal the southward shift of the ITCZ, that is an anomalous southward wind with decreasing NH average $SST$, not represented by the local regressions.
Also the anomalous anticyclonic circulation is reproduced in the large--scale regressions by the dipoles over northeastern Atlantic (positive to the south and to the east).
The impact on AMOC stability of wind--stress feedbacks has been investigated in the recent paper by~\cite{Arzel2010}, where a simple zonally averaged atmospheric model was used.
Even though it is quite difficult to compare their results with the results from a GCM like SPEEDO, the general picture is similar.
The atmospheric circulation in the NH is strengthened, while the opposite is true for the SH.
The magnitude of the changes in SPEEDO is close to their lowest estimates.

\section{HCM test}
\label{sec:test}

The HCM consists of the ocean component of SPEEDO (i.e., CLIO) and the dynamic boundary conditions described in the previous section.
It was tested by comparing its results with the original SPEEDO model.
The first experiment (regCLIM) starts from the end state of the ocean of the CLIM run. 
The model is forced only by the local regressions (values of $p_2$ set to zero) for 3000 years.
Next, all the large--scale regressions are also switched on, and the model runs for 2000 years more.

\begin{figure}[tbp]
  \begin{center}
    \includegraphics[width=0.8\columnwidth]{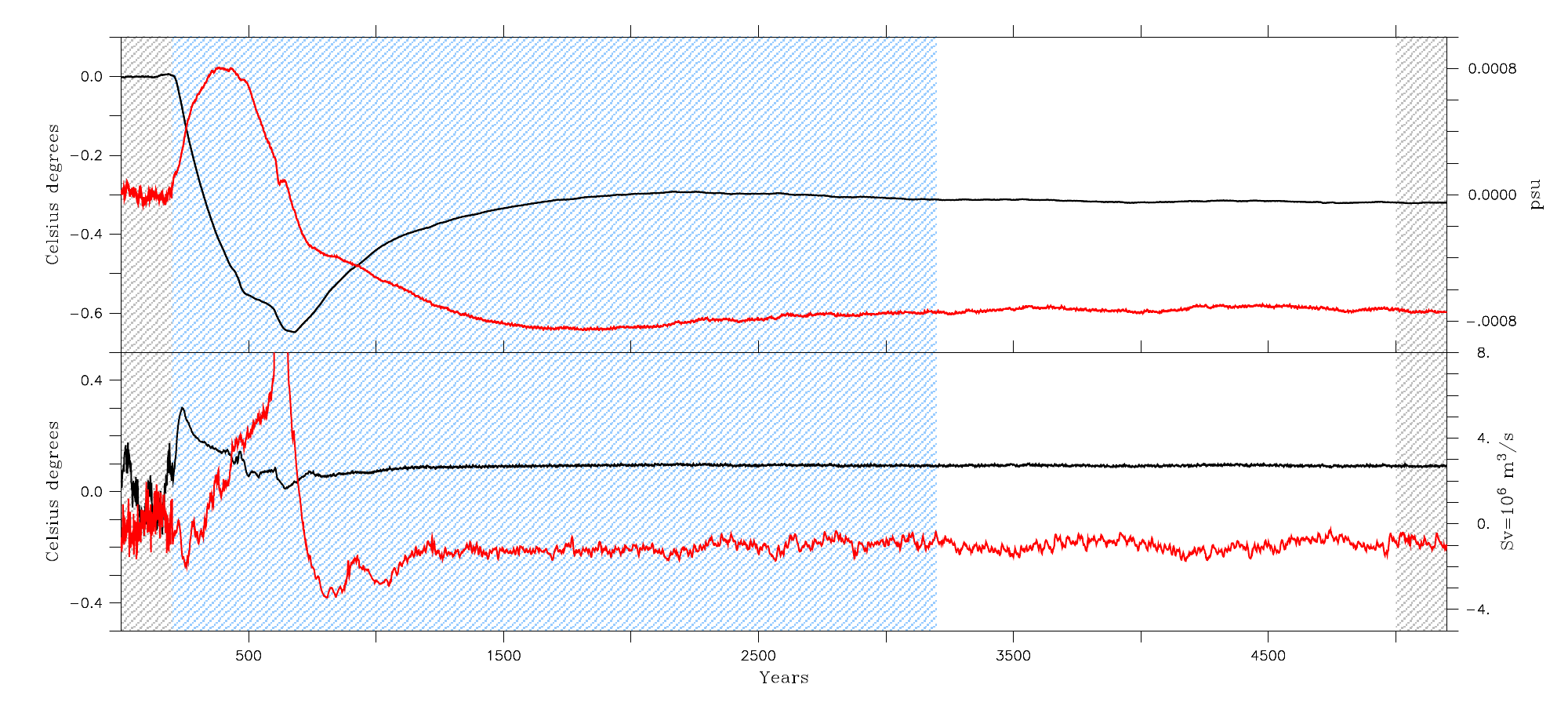}
  \end{center}
  \caption{Deviation from CLIM average in regCLIM of four quantities: global average sea temperature (top panel, black, left $y$--axis), global average salinity (top panel, red, right $y$--axis), global average $SST$ (bottom panel, black, left $y$-axis) and maximum AMOC (lower panel, red, right $y$--axis).}
  \label{fig:model_drift}
\end{figure}

\begin{figure}[tbp]
  \begin{center}
    \includegraphics[width=0.8\columnwidth]{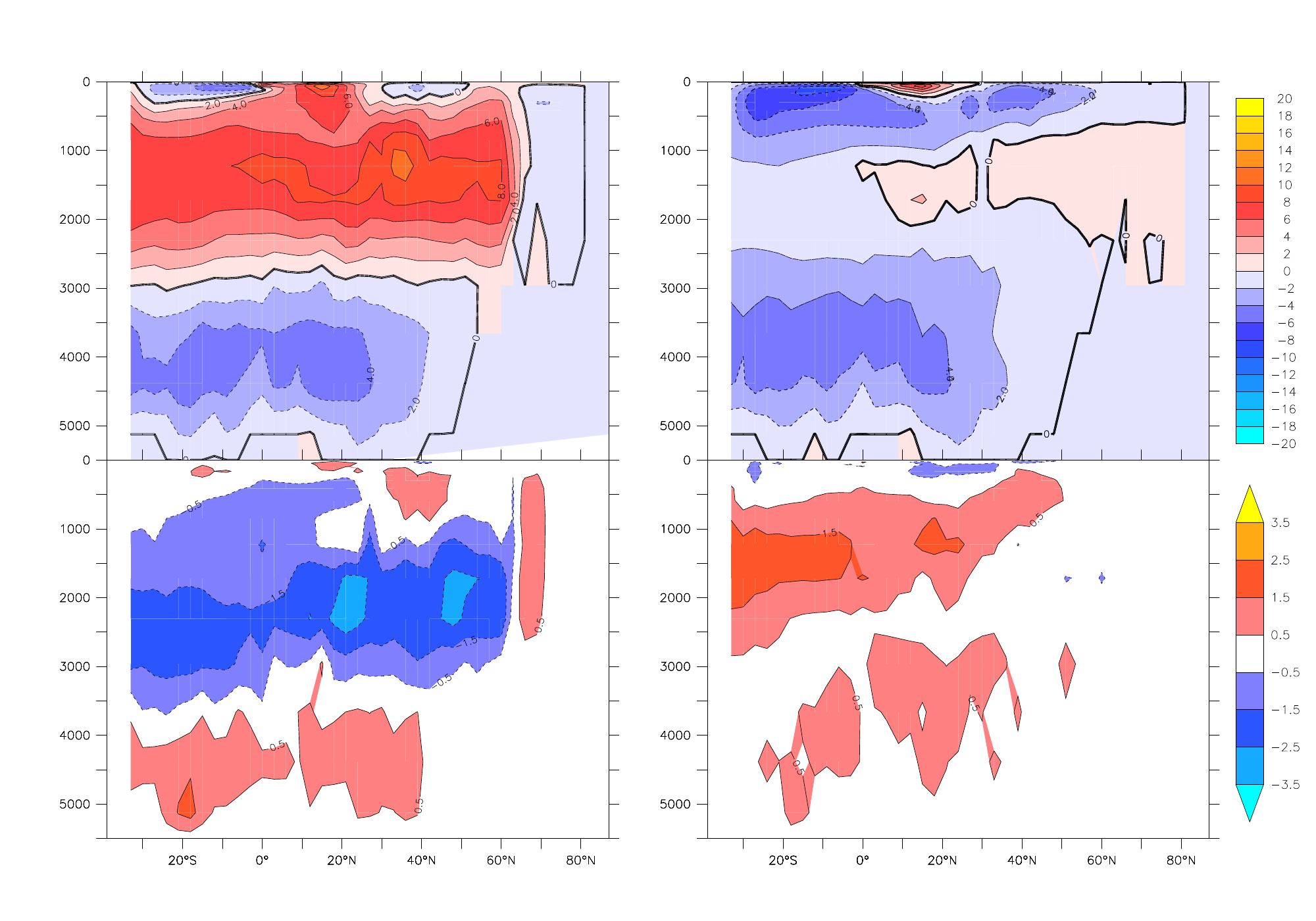}
  \end{center}
  \caption{Overview of overturning streamfunction in the various models. 
  In the top panels, AMOC for the CLIM mean state (top left) and for years  101 to 110 of PULSE (top right) are shown. 
  The shaded contours are every $2\;\mathrm{Sv}$, the red filling is for positive values, blue for negative.
  The thick line is the zero contour.
  In the left bottom panel, the difference of the overturning streamfunction between the last 200 years of regCLIM and the CLIM mean state is shown. 
  In the right bottom panel, the difference of the overturning streamfunction between years 101 to 110 of regPULSE and PULSE runs.
  The contours in the lower panels are every $1\;\mathrm{Sv}$.
  }
  \label{fig:Psia}
\end{figure}

Results of the regCLIM run are shown in figure~\ref{fig:model_drift}. 
On the top panel, the deviation from CLIM mean value of the global average sea temperature (salinity) is reported in black (red). 
The area shaded in grey on the left margin of figure~\ref{fig:model_drift} marks the (200 years) data of the CLIM run. 
The light blue area marks the first 3000 years of the regCLIM of the ocean--only model, with only local regressions active.
To estimate the theoretical equilibrium state of the model, we fit the global average sea temperature and salinity from years 1201--5200 of regCLIM with the function:
\begin{equation}
  f(t) = a_1 \mathrm{sin}\left (\frac{t+a_2}{a_3} \right ) \mathrm{exp}\left [ - \frac{t+a_2}{a_4} \right ] + B,
  \label{eq:fit_func}
\end{equation}
where $t$ is time, $a_1, \ldots, a_4$ are the fit parameters, and $B$ is a constant background that represents the state of the system at infinite time. 
The theoretical equilibrium state computed from this procedure is $0.31 ^\circ \mathrm{C}$ colder and $7.2 \cdot 10^{-4} psu$ fresher than the coupled CLIM run.
Little drift, but a substantial reduction of the variability due to the restoring term, is observed in the global average $SST$ (figure~\ref{fig:model_drift}, black line in the bottom panel).
The NH average $SST$ increases by $0.18 ^\circ \mathrm{C}$ (difference between last 200 years of regCLIM and CLIM).
The maximum of the AMOC is, at the end of regCLIM, approximately $1\;\mathrm{Sv}$ weaker than in the CLIM run (bottom panel of figure~\ref{fig:model_drift}, in red).
The AMOC, as the left bottom panel of figure~\ref{fig:Psia} shows, is weaker and approximately $500 \mathrm{m}$ shallower in the HCM.
The freshwater transport by the AMOC at $30^\circ\mathrm{S}$ in the last 200 years of regCLIM (grey shaded area on the right of figure~\ref{fig:model_drift}) is $M_{ov} = -0.06\;\mathrm{Sv}$.
To keep $M_{ov}<0$, the freshwater corrections described in section~\ref{sec:Model} are 50\% stronger than in the fully coupled model.

\begin{figure}[tbp]
  \begin{center}
    \includegraphics[width=0.8\columnwidth]{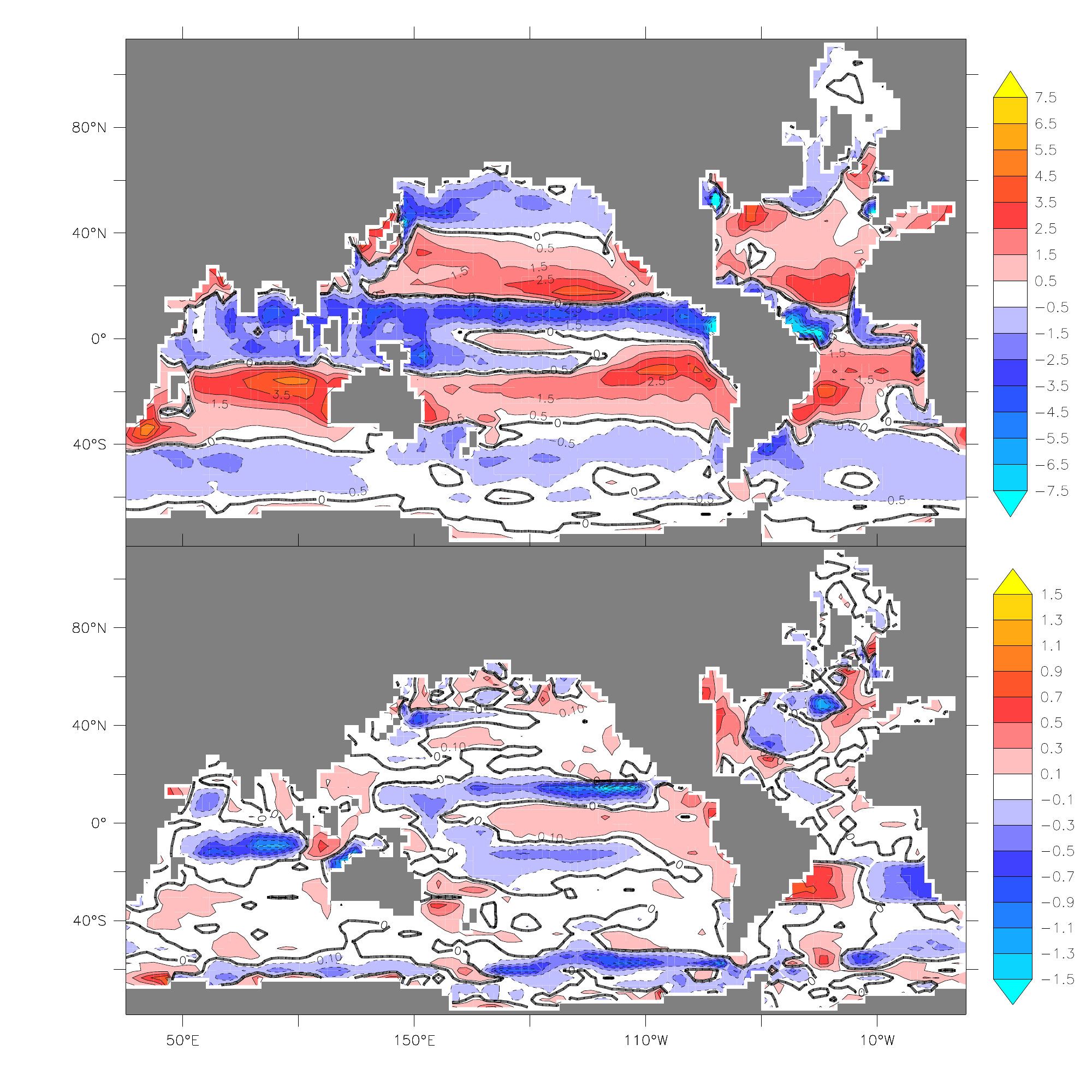}
  \end{center}
  \caption{In the top panel, the surface density flux for CLIM is shown in $10^{-6} \cdot kg/(m^2 \; s)$. 
  In the bottom panel, the difference of the same quantity between the last 200 years of regCLIM and CLIM. 
  Different colour scales are used in the two panels.
  In the figure, the grid of the ocean model is used (distorted in the north Atlantic and Arctic), to avoid interpolation errors.
  }
  \label{fig:rhoflux}
\end{figure}

To investigate the origin of the changes in the AMOC strength, we diagnose the surface fluxes of density for the CLIM and regCLIM runs.
The surface density flux $\Phi_\rho$ can be estimated using the formula~\cite[]{Gulev2003,Tziperman1986}:
\begin{equation}
  \label{eq:rhoflux}
  \Phi_\rho = -\frac{\alpha}{c_p}\Phi_H + \rho_0 \beta \frac{\Phi_E \cdot SSS}{1-SSS \cdot 10^{-3}},
\end{equation}
where $\alpha= - 1/\rho_0\left(\partial \rho / \partial T\right)$, $\beta=1/\rho_0\left(\partial \rho / \partial S\right)$, $\Phi_H$ is the total surface heat flux into the ocean ($\Phi_H = \Phi_Q+\Phi_{SW}$), $\rho_0$ is the reference water density, $SSS$ is the surface salinity measured in ppt. 
The density flux into the ocean is shown in figure~\ref{fig:rhoflux} in units of $10^{-6} \cdot kg/(m^2 \;s)$ for the CLIM run (top panel).
The effect of sea--ice cover is taken into account in the computation of the density flux, and the model grid (distorted over north Atlantic and Arctic) is used to avoid interpolation errors.
The difference between the fluxes from the regressions in the last 200 years of regCLIM and CLIM is reported in the bottom panel of figure~\ref{fig:rhoflux}.
Even if the changes are generally small (note the different colour scales in the figure), when the difference is averaged over the GIN seas and the Arctic Mediterranean (taking as southern boundaries the Bering strait and the latitude of the southern tip of Greenland), we find that the density flux decreases by $2\cdot 10^{-8} kg/(m^2 \; s)$.
This value represents a 10\% decrease of the average density flux over the same area, that nicely fits the relative change in maximum overturning strength.

\begin{figure}[tbp]
  \begin{center}
	  \includegraphics[angle=270,width=0.8\columnwidth]{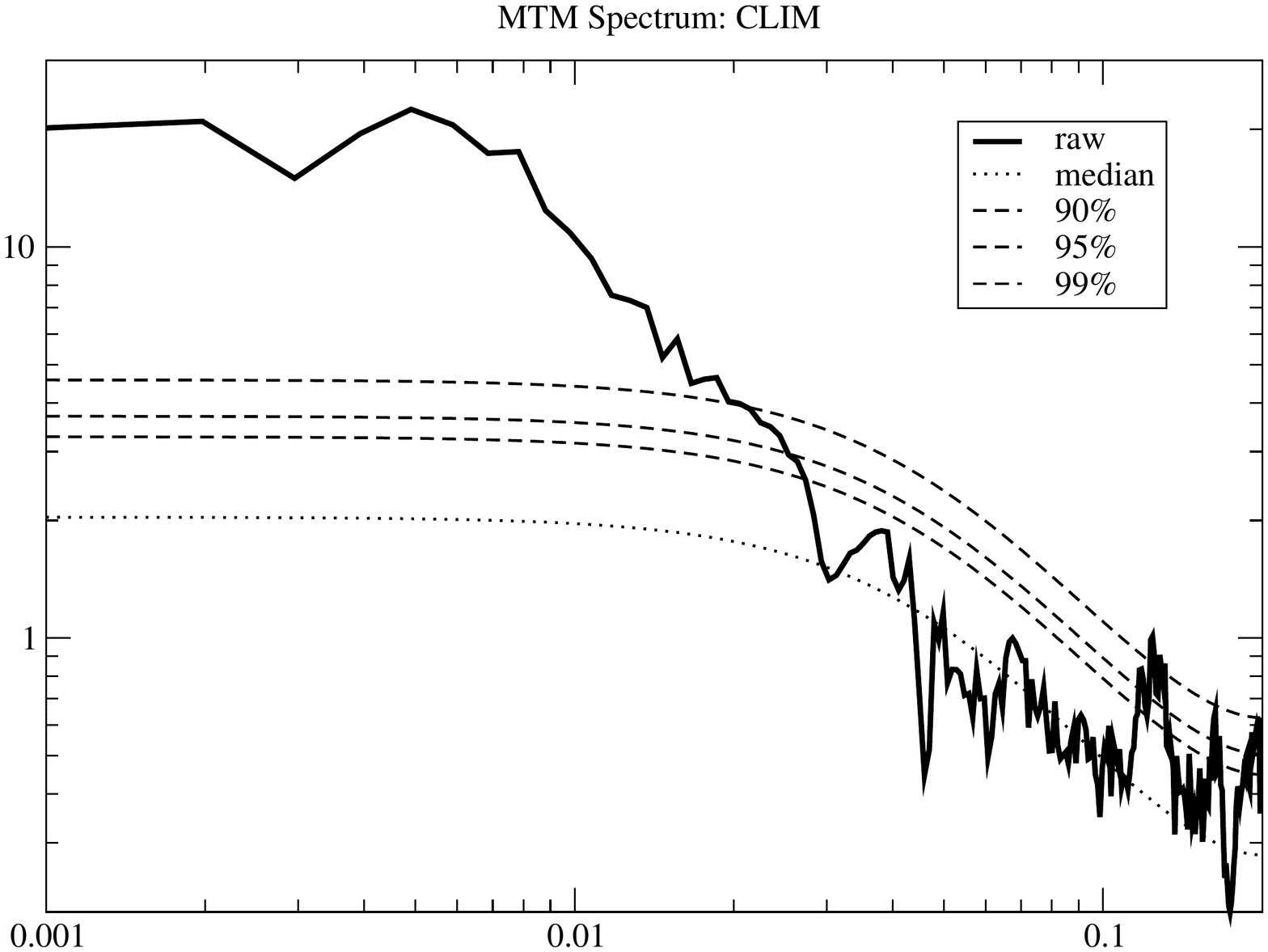}

	  \includegraphics[angle=270,width=0.8\columnwidth]{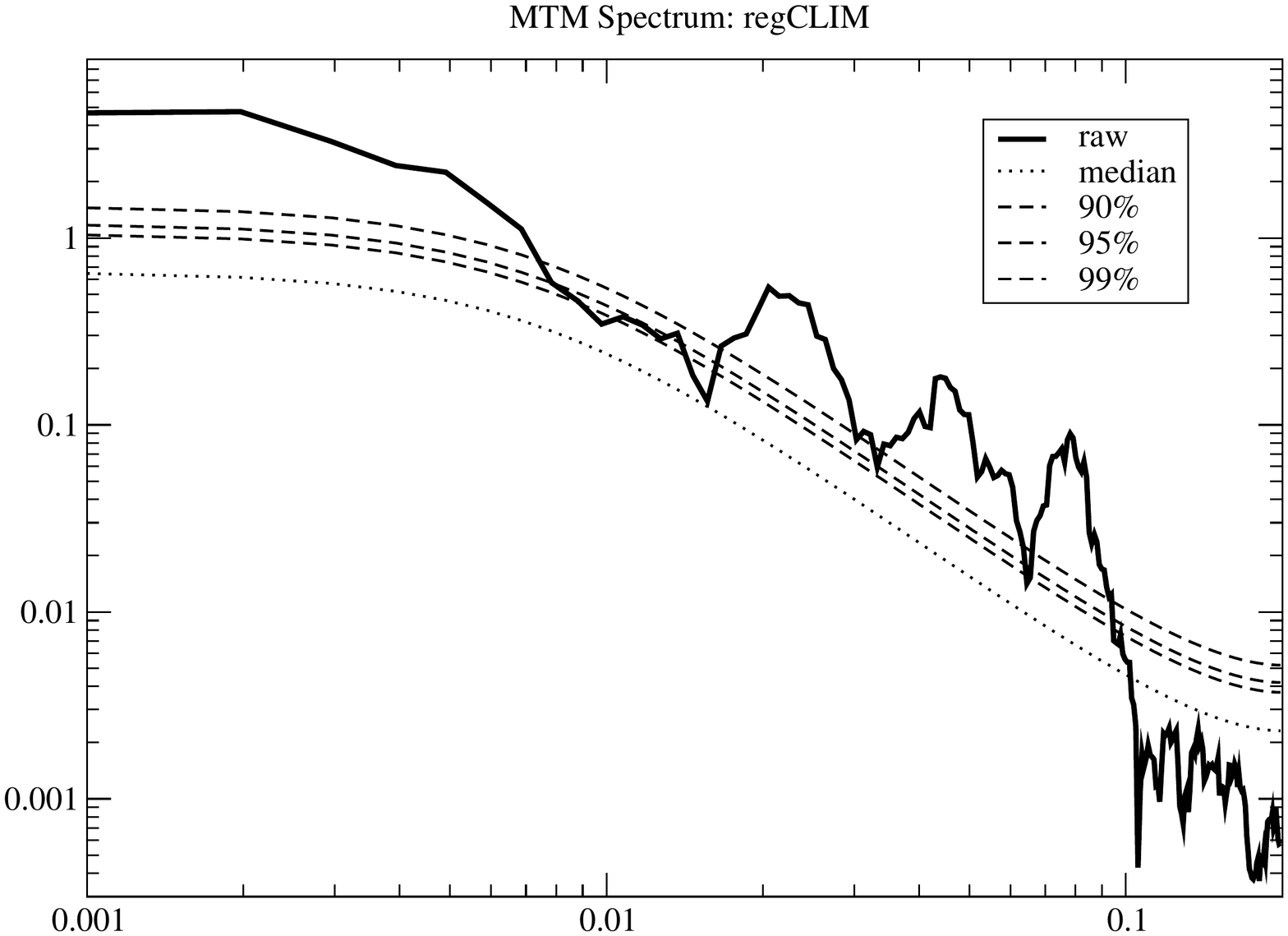}
  \end{center}
  \caption{MTM spectra of the time series of the maximum AMOC (solid lines) for CLIM (top panel) and regCLIM (bottom panel).
  The dashed smooth lines represent, from the lowest to the highest, the estimated red noise background and the median, 90\%, 95\% and 99\% significance levels associated with it.
In both cases, the resolution is $(5 years)^{-1}$ and 7 tapers were used.
Time series are 1000 years long.
  }
  \label{fig:spectra}
\end{figure}

The definition of the HCM, that does not include any high frequency stochastic component, causes a strong reduction of variability, but low frequency variability of the system seems to be preserved.
To show this, a multi taper method (MTM) analysis \cite[]{Ghil2002} was performed on the time series of the maximum of overturning streamfunction of the Atlantic.
The analysis is performed on the yearly data of CLIM (a longer control run is used, 1000 years long) and the last 1000 years of regCLIM (figure~\ref{fig:spectra}).
At the lower end of the spectrum, energy is concentrated at similar frequencies in the two models, below approximately $0.02 \; \mathrm{year} ^{-1}$.
At higher frequencies, instead, the broad peaks found in the HCM between $0.02\;\mathrm{year}^{-1}$ and $0.09\;\mathrm{year}^{-1}$ are not present in the original coupled model, while the peaks found above $0.1\;\mathrm{year}^{-1}$ in CLIM are lost in the HCM.
Also the first empirical orthogonal function of $SST$ computed from the HCM resembles the one from CLIM only in the northwestern Atlantic.
This approach is thus limited when the internal variability of the ocean is of interest, but in the present work the focus is only on the quasi--equilibrium response.
Atmospheric noise and lagged correlations are probably needed to better represent and excite the modes of variability of the system.

As a final test, a pulse experiment was performed in the HCM.
In this test, that will be shorthanded as regPULSE, we apply the same freshwater anomaly as in PULSE (see section~\ref{sec:regressions}), also increased by 50\% as the corrections already applied in regCLIM.
The initial conditions for regPULSE are provided by the final state of regCLIM: year 5200 of figure~\ref{fig:model_drift}.
In regPULSE, as in PULSE, the anomaly is applied for 1000 years, letting the model reach a new equilibrium afterwards. 
We focus our analysis on the response of the system during the first hundred years of the run, where the regressions are expected to be significant.

\begin{figure}[tbp]
  \begin{center}
    \includegraphics[width=0.8\columnwidth]{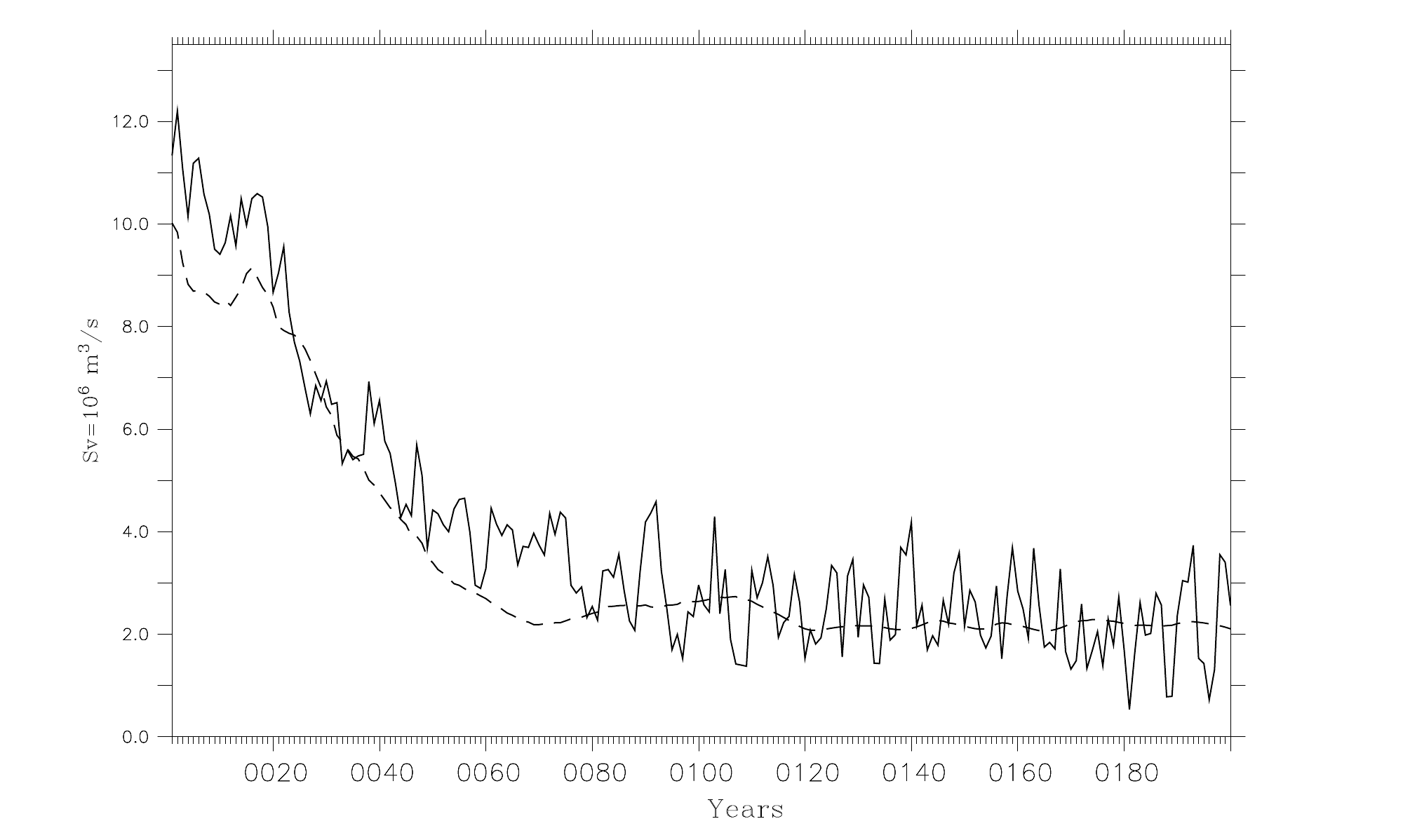}
  \end{center}
  \caption{
  Maximum AMOC for first 200 years of PULSE (full line) and regPULSE (dashed line) in $\mathrm{Sv}$.
  }
  \label{fig:streams_reg}
\end{figure}

The AMOC maximum for regPULSE is reported in figure~\ref{fig:streams_reg} as a dashed line.
The response of the AMOC in regPULSE, when measured by this quantity, follows closely the one in PULSE.
The only substantial differences are its lower initial condition and the weaker variability of the regPULSE signal.
The weaker variability of regPULSE signal is no surprise, considering the fact that our regressions do not add any high frequency variability to the system, depending only on $SST$.

Looking at the entire overturning streamfunction of the Atlantic, the results are also encouraging.
On the right hand side of figure~\ref{fig:Psia}, the overturning of the collapsed state that is established after the first 100 years of the pulse experiment are compared in PULSE and regPULSE.
In the top right panel of figure~\ref{fig:Psia}, the streamfunction of years 101--110 of PULSE run is shown as a reference.
The difference between regPULSE and PULSE during the same years is reported below.
The results of the HCM are in good agreement with those of SPEEDO, showing a reversed cell only slightly weaker than in PULSE.
The largest differences are at the southern border of the Atlantic basin, likely in connection with the general underestimation of the density flux over the southern ocean (figure~\ref{fig:rhoflux}).
For what concerns the barotropic streamfunction during the pulse experiments, the only significant differences are found in the southern ocean (not shown).
Over the Pacific sector of the Southern ocean, the underestimation of the barotropic streamfunction represents about 20\% of the transport predicted by PULSE.
This discrepancy is probably connected with an overestimation of the decrease of the southern westerly winds in the regressed forcing in response to the collapse of the AMOC.

\section{Summary and Conclusions}\label{sec:Conclusions}

In this paper we described a new technique for developing a global HCM that includes a basic representation of the feedbacks due to the ocean--atmosphere interaction, relevant for the stability of the AMOC.
The steady state feedbacks of the system were represented through linear regression terms depending on the local deviation of $SST$ from its mean value. 
The large--scale response of the atmosphere to an externally forced AMOC collapse is included with a regression on the NH hemisphere average temperature.

The results of the regressions give a quantitative representation of the changes in the surface fluxes that is consistent with other model experiments~\cite[]{Stouffer2006,Krebs2007,Laurian2009}.
In particular, we can detect the changes in heat flux at the surface due to the cooling of the NH after a AMOC collapse. 
Significant changes are observed also in the freshwater flux, in connection with the response of the general circulation in the atmosphere to the changes in the equator to pole temperature profile, that determine the response of the winds as well.
The boundary conditions computed in section~\ref{sec:regressions}, were then successfully used as a dynamic forcing for an ocean--only model.

This ocean forced by a ``minimal atmospheric model'' guarantees a decrease of the computation time between ten and twenty times with respect to the original coupled model.
The ocean model forced by the regressions which form the HCM reaches a steady state close to the one of the original coupled model.
Furthermore, an experiment is performed where the AMOC is collapsed in both the fully coupled model and in the ocean forced by the regressions.
The two results are in good agreement.
This enables us to proceed to further use the HCM to investigate the impact of the atmospheric feedbacks on the stability of the AMOC.
In particular, the formulation of the forcing shown in section~\ref{sec:regressions} enables us to selectively choose which fluxes are fixed to a climatological value, and which ones are computed dynamically as a function of $SST$.
We can thus investigate the impact of each feedback separately on quantitative grounds, and we can aim at a deeper understanding of the main physical processes involved in the collapse and recovery of the AMOC.
It is also important to analyse the response of the HCM to weaker freshwater anomalies.
Reducing the anomaly that forces the AMOC collapse, the atmospheric feedbacks are likely to play an increasingly dominant role.

The model can obviously be extended in many ways.
Using higher order (nonlinear) models in the data fit is unlikely to be worth the effort.
The study of the role of atmospheric noise and of correlations lagged in space and time, and their inclusion in the HCM, may instead greatly improve the representation of atmosphere--ocean interaction with respect to the variability of the AMOC.

As a final remark, we want to stress that our technique to design the HCM is general.
We do not rely on any ad--hoc assumption connected with the nature of the EMIC that was used for this work. 
For this reason, this technique is potentially interesting for many other problems (apart from the stability of the AMOC) where a computationally efficient, simple representation of the ocean--atmosphere interaction is desired.
For instance, instead of using data from the atmospheric component of SPEEDO, the ocean component could be coupled to a statistical atmosphere derived from a state--of--the--art coupled climate model or from reanalysis data, at least for the computation of local regressions.

\begin{acknowledgments}
This work is funded by the Netherlands Organisation for Scientific Research. We acknowledge Camiel Severijns (KNMI) for his precious technical support, and Matthijs den Toom (IMAU) for the stimulating discussions.
\end{acknowledgments}

  \bibliography{Article1}

  \clearpage

\begin{table}[tbp]
  \centering
  \begin{tabular}{p{2.3cm} | p{1.5cm} p{1.8cm} p{1.8cm}}
    Name & Model & Freshwater anomaly & Total length (years)\\
    \hline
    CLIM & EMIC & None & 200 \\
    PULSE & EMIC & $0.4\;Sv$  & 4000\\
    regCLIM & HCM & None & 5000 \\
    regPULSE & HCM & $0.6\;Sv$ & 5000 \\
  \end{tabular}
  \caption{List of the model runs described in the text.}
  \label{tab:runs}
\end{table}

\end{document}